\documentclass[twocolumn,showpacs,preprintnumbers,amsmath,amssymb]{revtex4}
\usepackage[english]{babel}
\usepackage{graphicx}% Include figure files
\usepackage{dcolumn}% Align table columns on decimal point
\usepackage{bm}

\begin{document}

\title{Comments on "Stability of Tsallis entropy and instabilities of Renyi and\\normalized Tsallis entropies:
A basis for $q$-exponential distributions"}% Force line breaks with \\

\author{A. G. Bashkirov}
\email{abas@idg.chph.ras.ru}

\affiliation{Institute Dynamics of Geospheres, Russian Academy of
Sciences, Moscow, Russia} %Lines break automatically or can be forced with \\

\date{\today}% It is always \today, today,
\pacs{65.40Gr, 05.20.-y, 05.90.+m, 02.50.-r}
\maketitle

Abe \cite{Abe} presented two counterexamples of instability of the
Renyi entropy% following the paper by Leshe \cite{Le}
and showed that the Tsallis entropy is stable for these
counterexamples. From the time of its publication this work is
often referred (see e.g. \cite{TsJMP,Ka}) as a mortal verdict for
the Renyi entropy. On the other hand, the Renyi entropy is widely
used now. Because of this, the main points of Ref. \cite{Abe}are
to be revised carefully.

Abe calculated responses $\left|\Delta S^{(R)}\right|$ and
$\left|\Delta S^{(Ts)}\right|$ to small variation of initial model
distributions over $W$ states of a system and then passed to the
limit $W\to\infty$ treating an amplitude $\delta$ of the variation
as a finite constant. As a result, he found a loss of continuity
of a response of the Renyi entropy to the small perturbations. On
my point of view, such a conclusion can be dismissed on two
counts. First, Abe considered normalized values of $\left|\Delta
S^{(R)}\right|$ and $\left|\Delta S^{(Ts)}\right|$ with different
$W$-dependent normalization factors, $S^{(R)}_{max}$ and
$S^{(Ts)}_{max}$, correspondingly. Such the normalization
influenced on their limiting properties and, consequently, on
conclusions about their stabilities. Second, continuity is to be
checked with the use of the opposite iterated limiting process:
firstly, $\delta\to 0$ and then $W\to\infty$. Such an order
corresponds to a traditional approach in statistical physics where
all properties are calculated firstly for finite systems and the
thermodynamic limit is performed after all calculations (see, e.
g. \cite{Isih}). Below are modifications of Abe's results for such
order of the limiting procedures.

For brevity sake, the first of Abe's counterexamples alone will be
discussed here. It is especially important, because of it refers
to the range $0<q<1$ of the most if not all of applications
\cite{Bash} of the Renyi entropy. The second counterexample may be
discussed in the same manner.

The examined small ($\delta\ll 1$) deformation of distribution
$\{p\}$  over $W$ states ($W\gg 1)$ for $0<q<1$ are
\begin{equation}
p_i=\delta_{i1},\,\,\,p_i'=\left(1-\frac \delta{2}\frac
W{W-1}\right)p_i+\frac \delta{2}\frac 1{W-1}.
\end{equation}

Using the well-known definitions of the Tsallis and Renyi
entropies (for $k_B=1$), we get
%\begin{widetext}
\begin{eqnarray}
\left|\Delta S^{(Ts)}\right|&=&\frac 1{1-q}\left[\left(1-\frac
\delta{2}\right)^q\right.\nonumber\\
&+&\left(\frac \delta{2}\right)^q(W-1)^{1-q}-1\biggr]
%&\simeq&\frac 1{1-q}\left[-q\frac \delta{2}+\left(\frac
%\delta{2}\right)^q(W-1)^{1-q}\right].%,\,\,\,\,0<q<1,
\end{eqnarray}
%On the other hand,
\begin{eqnarray}
\left|\Delta S^{(R)}\right|&=&\frac 1{1-q}\ln\left[\left(1-\frac
\delta{2}\right)^q+\left(\frac
\delta{2}\right)^q(W-1)^{1-q}\right]\nonumber\\
&<&\left|\Delta S^{(Ts)}\right|,%\,\,\,\,0<q<1,
\end{eqnarray}
%\end{widetext}
 where the last inequality is resulted from the fact
that the logarithm as a concave function is always less than its
linearized approximation. Thus, stability of the Renyi entropy for
the counterexample (1) is at least not lower stability of the
Tsallis entropy.

I may suppose that Abe paid no attention to this evident
inequality  because of he considered their normalized values with
different $W$-dependent normalization factors. Such the
normalization influenced on their limiting properties and,
consequently, on conclusions about their stabilities. It seems
more reasonable to normalize the gains with the number of states
$W$. In this case, $\left|\Delta S\right|/W$ is the entropy gain
per a state and the limit $W\to\infty$ corresponds to the
thermodynamic limit in statistical physics. It is evident from the
above equations that the gain per a state for each of the
discussed entropies tends to zero when $W\to\infty$.

As for stability of the Tsallis entropy, there is no double limit
($\delta\to 0$, $W\to\infty$) of $\left|\Delta S^{(Ts)}\right|$ as
a function  of $\delta$ and $W$ but there is a repeated limit
($\delta\to 0$ and then $W\to\infty$) and it is equal to zero.
Indeed, this function  becomes infinitesimal for any finite $W$
when
\begin{equation}
\delta/2\ll (W-1)^{-\frac{|1-q|}{q}}.
\end{equation}

%{99}

\end{document}